\documentclass[aps,prl,twocolumn]{revtex4}
\usepackage{amssymb}
\usepackage{amsfonts}
\usepackage{amsmath}
\usepackage{graphicx}

\begin{document}

\title{A Strongly Anisotropic Superconducting Gap in the Kagome Superconductor CsV$_3$Sb$_5$: A Study of Directional Point-Contact Andreev Reflection Spectroscopy}

\author{Yu-qing Zhao$^{1\S}$, Zhi-fan Wu$^{1\S}$, Hai-yan Zuo$^1$, Weiming Lao$^1$, Wangju Yang$^1$, Qiuxia Chen$^1$, Yao He$^1$, Hai Wang$^1$, Qiangwei Yin$^2$, Qi Wang$^3$, Yang-peng Qi$^{3}$, Gang Mu$^4$, He-chang Lei$^{2\dag}$, and Cong Ren$^{1,5\ddag}$}

\address{$^1$ School of Physics and Astronomy, Yunnan University, Kunming 650500, China}
\address{$^2$ Department of Physics and Beijing Key Laboratory of Opto-electronic Functional Materials and Micro-nano Devices, Renmin University of China, Beijing 100872, China}
\address{$^3$ School of Physical Science and Technology, ShanghaiTech University, Shanghai 201210, China}
\address{$^4$ State Key Laboratory of Materials for Integrated Circuits, Shanghai Institute of Microsystem and Information Technology, Chinese Academy of Sciences, Shanghai 200050, China}
\address{$^4$ Yunnan Key Laboratory for Electromagnetic Materials and Devices, Yunnan University, Kunming {\rm 650500}, China}


\begin{abstract}
In the recently discovered V-based kagome superconductors AV$_3$Sb$_5$ (A = K, Rb, and Cs), superconductivity is intertwined with an unconventional charge density wave (CDW) order, raising a fundamental concern on the superconducting gap structure of such kagome superconductor in the presence of CDW orders. Here, we report directional soft point-contact Andreev reflection (SPCAR) spectroscopy measurements on the kagome superconductor CsV$_3$Sb$_5$, revealing compelling evidence for the existence of a strongly anisotropic superconducting gap pairing state.  The SPCAR spectra measured with current injected parallel to the $ab$-plane exhibit an in-gap single conductance peak, in contrast to those of SPCAR spectra: a double-peak structure in the perpendicular direction. These spectra are well described by an anisotropic single-gap BTK model. The extracted superconducting gaps comprise an isotropic large gap and a strongly anisotropic gap, originating from different Fermi surface sheets. Quantitative analysis reveals an anisotropy around $\sim$ 70\% with a gap minimum of about 0.15 meV. These results shew new light on the unconventional multiband pairing states in kagome superconductors.
\end{abstract}

\maketitle

\emph{1.Introduction:} Kagome lattices, characterized by their unique two-dimensional network of corner-sharing triangles, have long captivated the condensed matter physics community as fertile ground for exploring novel quantum phenomena, which arise from strong electron correlations and magnetic frustration inherent to the kagome geometry \cite{a1,a2,a3,a4,a5,WangQH2013}.  This paradigm shifted dramatically with the recent discovery of superconductivity in a new class of materials: the vanadium-based kagome superconductors (SCs) AV$_3$Sb$_5$ (A = K, Rb, Cs). These compounds realize an ideal, non-magnetic kagome net of vanadium atoms with superconducting critical temperatures $T_c\sim 1-3$ K \cite{WenXG2009,LiJX2012,Ortiz2020,LeiCPL2021,Ortiz2019}. More intriguingly, the emergence of superconductivity is intertwined with unconventional charge density wave (CDW) orders in these kagome metals \cite{a9,a10,a10a,Tan2021,Nie2022,Ortiz2024,HuJP2023,a13,a14,a15}.  Understanding the nature of this superconductivity and its interplay with the CDW orders within the unique kagome framework, potentially distinct from conventional mechanisms, is now a central focus in quantum materials research.

\begin{figure*}
\includegraphics[scale=0.55]{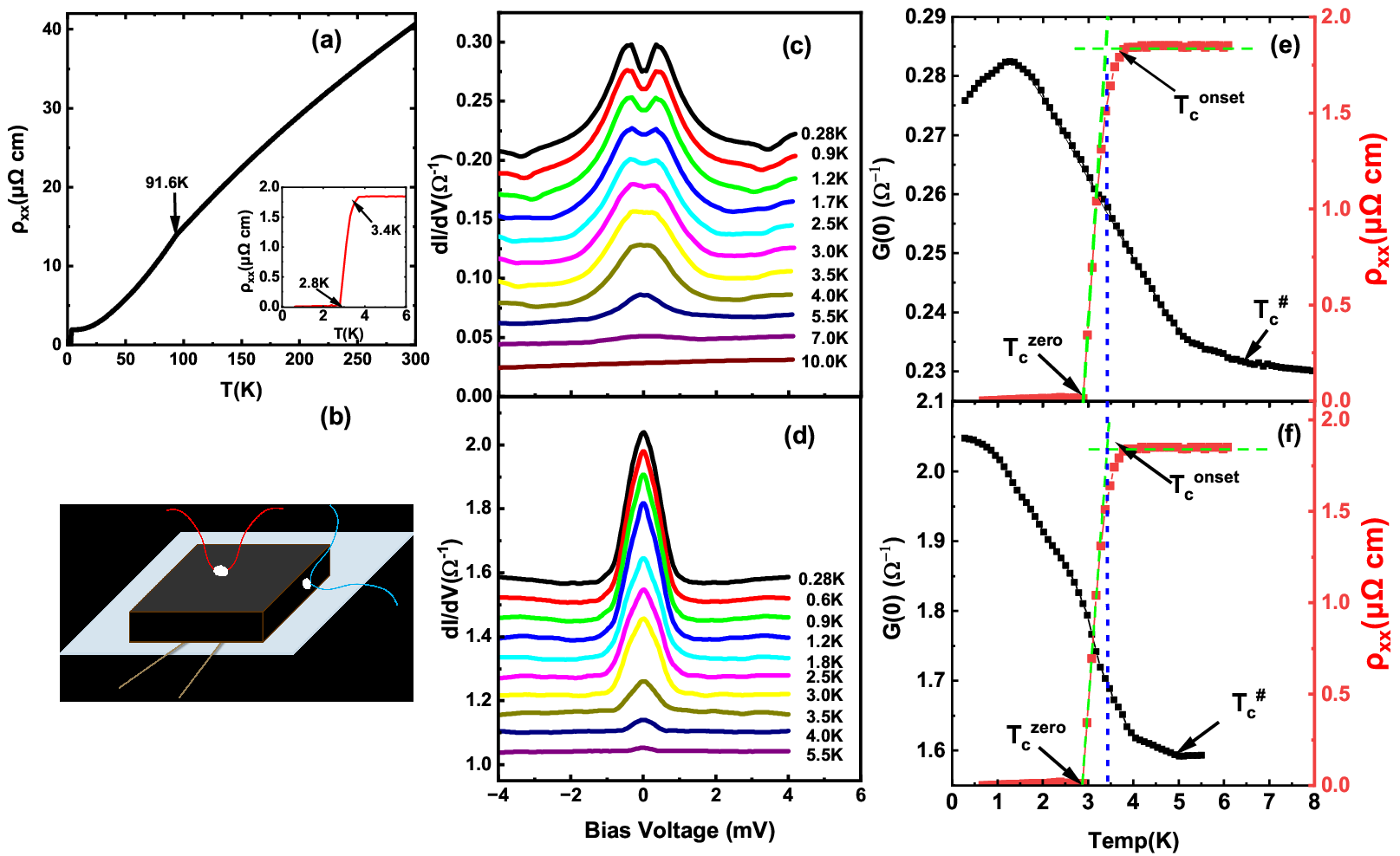}
\caption{(a) Temperature dependence of the in-plane resistivity $\rho_{xx}$ of CsV$_3$Sb$_5$ single crystal, a pronounced kink at approximately 91.6 K indicates the onset of the charge density wave (CDW) transition. The inset magnifies the low-temperature region, showing the superconducting transition with the onset temperature $T_c^{\mathrm{onset}}$ and zero-resistivity temperature $T_c^{\mathrm{zero}}$ marked. (b) Schematic of the ``soft'' point-contact spectroscopy measurement configuration. The black block represents the CsV$_3$Sb$_5$ single crystal, while the red and blue lines indicate the injection paths for soft point-contact Andreev reflection measurements with current applied perpendicular (labeled as $c$-axis junction) and parallel (labeled as $ab$-plane junction) to the $ab$-plane, respectively. (c)(d) $T$-dependent point-contact conductance spectra $G\equiv dI/dV$ measured with sensing current $I$ along $c$-axis (``$c$-axis'' junction) (c) and along $ab$-plane ($ab$-plane junction), respectively. (e)(f) $T$-dependence of the zero-bias conductance $G(0)$ of CsV$_3$Sb$_5$/Ag point contacts for $c$-axis (e) and $ab$-plane junctions (f), respectively.  The bulk resistivity $\rho_{xx}$ in the superconducting transition region is presented for comparison.}
\end{figure*}

As a fundamental parameter clarifying the microscopic pairing mechanism and interplays between multiple electronic orders, the superconducting gap structure or order parameter in kagome superconductors remain highly controversial and no consensus has yet been reached.  Indeed, the observations of the Hebel-Slichter coherent peak in the spin-lattice relaxation rate from the NMR studies of CsV$_3$Sb$_5$ indicated a nodeless $s$-wave superconductivity \cite{NMRCPL2021}.  Magnetic penetration depth \cite{YuanHQCPL2021} along with specific heat \cite{a10a} measurements collectively point to two nodeless gaps in weak-coupling limit in CsV$_3$Sb$_5$.  Such two isotropic gaps but in strong coupling limit have been confirmed in point-contact Andreev reflection spectrum \cite{LuXPRB2021,HeMCPRB2022}.  On the contrary, thermal conductivity measurements in CsV$_3$Sb$_5$ and $\mu$SR measurements in K/RbV$_3$Sb$_5$ have suggested the presence of a nodal gap in such V-based kagome superconductors \cite{LiSY,LeiHCnpj2022}.  While low-$T$ scanning tunneling microscopy (STM) appeared to show both nodal and nodeless sign-preserving gaps with multiple Fermi surfaces for the same material \cite{FengDLPRL2021,Chen2021}, the latest angle-resolved photoemission spectroscopy (ARPES) measurements of high-quality CsV$_3$Sb$_5$ single crystals revealed that the SC gap in the $\beta$ Fermi surface is strongly anisotropic with anisotropy $\sim$80\%, suggesting the possible presence of a node gap \cite{lateARPES2024}.  Moreover, it has been reported that the anisotropy in the SC gap of CsV$_3$Sb$_5$ can be suppressed by electron irradiation \cite{elecrad}, leading to a nodal-nodeless gap structure transition, similar to the case of nodal $s_{\pm}$-wave SC BaFe$_2$(As$_{1-x}$P$_{x}$)$_2$ \cite{impuritytheory,Matsuda1,Matsuda2}. These studies with different conclusions reflect the complexity of the system, and the direct measurement of the SC gap structure in such kagome superconductor is highly desired.

Point-contact Andreev reflection (PCAR) spectroscopy, some extent equivalent to tunneling spectroscopy, is a well-established technique for probing gap structure of superconductor \cite{Yanson,Gonnelli}.  Based on some sophisticated theoretical models, PCAR spectra have also been proven to be suitable to study unconventional superconductors with ``exotic'' properties such as anisotropic, nodal/nodeless, and multiple SC gaps. One advantage of PCAR spectroscopy in studying the superconducting gap structure is that point-contact junction can be flexibly fabricated on the surface of the superconducting crystals under study.  In this study, we conduct a directional PCAR spectroscopy study on CsV$_3$Sb$_5$ to investigate its superconducting gap structure by fabricating $c$-axis and $ab$-plane point contacts on CsV$_3$Sb$_5$ single crystals. These conductance spectra clearly feature the presence of an anisotropic gap besides two isotropic gaps.

\emph{2. Experiment Details:} Single crystals of CsV$_3$Sb$_5$ were synthesized using the self-flux method, in two different laboratories whose crystals are labeled as \#1 and \#2, respectively.  Figure 1(a) displays the temperature $T$ dependence of the in-plane resistivity, $\rho_{xx}(T)$, measured on a single crystal of CsV$_3$Sb$_5$. The crystal demonstrates a metallic characteristics, quantified by a residual resistivity ratio defined as $RRR \equiv R(300,\mathrm{K})/R(5,\mathrm{K}) = 22$. A resistivity anomaly around 92 K corresponds to a well-defined charge density wave (CDW) phase \cite{a10a}. The inset of Fig. 1(a) further details the superconducting transition observed at low temperatures, identifying a superconducting onset temperature $T_c^{\mathrm{onset}} \approx 3.4$ K and a zero-resistance temperature $T_c^{\mathrm{zero}} \approx 2.8$ K.

Point-contact to the CsV$_3$Sb$_5$ crystals were formed by a so-called ``soft'' point-contact technique \cite{Gonnelli2001}. This method involves attaching a platinum wire with a diameter of 16$\mu$m, coated with a 50–100$\mu$m in size of silver paint, to the cleaved surface of a CsV$_3$Sb$_5$ single crystal with a typical crystal dimensions $2.5\times 2.0\times 0.3-0.5$ mm$^3$ along the $a$, $b$, and $c$ crystallographic axes, respectively. In this configuration, the heterocontact is actually composed of many nanocontacts due to the nanocrystalline nature of the silver paint, mimic to the tip point-contact technique. Schematic of point-contact junction in our experiment are shown in Figure 1(b). Two kind of point-contact were formed: the write point-contact on $ab$ plane (labeled as ``$c$-axis'' junction) and the red-colored point-contact on $ac/bc$ plane (labeled as ``$ab$-plane'' junction).  For the $ab$-plane configuration, the crystal was cleaved perpendicular to the $ab$-plane and polished using 10,000-grit sandpaper to produce a flat surface ($ac/bc$-plane), onto which the platinum wire was then bonded with silver paint. The differential conductance spectra $G(V)$ were recorded with the standard phase-sensitive lock-in technique in a quasi-four-terminal configuration, where an ac modulation was applied to the sample on top of a DC current bias. The contact resistance $R_J$ between the nanoparticle of silver paint and CsV$_3$Sb$_5$ crystals was usually in the range of $0.5-10\ \Omega$, typical values of a genuine point contact between normal metals and superconductors.

\emph{3. Results and Discussions:} Figures 1(c) and 1(d) present the typical $T$-dependent $c$-axis and $ab$-plane SPCAR conductance spectra $G(V)\equiv dI/dV$ of CsV$_3$Sb$_5$ crystals, respectively. A striking feature in these two kind of $G(V)$ curves is that: at lowest $T$=0.28 K, a conductance a double peak around gap edge plus a zero-bias conductance dip in $c$-axis junction, on the other hand, an in-gap conductance peak in $G(V)$ for $ab$-plane junction.  Qualitatively, for highly transparent junctions at finite $T$s, the appearance of a double-peak structure at  gap-edge in the Andreev conductance spectrum is a characteristic of a nodeless gap state. In contrast, an in-gap conductance peak is a signature of an anisotropic gap state due to the presence of a finite DOS at low energy, like a $d$-wave gap in cuprates.  Another feature in these point-contact spectrum is the appearance of the excess conductance spectrum at $T>T_c$.  For example, Fig. 1(c) depicts $T$-evolution of $c$-axis point-contact spectra, exhibiting a two-distinct coherent peaks at gap-edge. As $T$ increases above $T_c\simeq 3.4$ K, significant residual Andreev conductance persists before completely disappearing above $T^{\#} \sim 6.5$ K. The same is true for $ab$-plane junctions shown in Fig. 1(d), that the $ab$-plane point-contact spectra vanish entirely at a temperature of $T^{\#} \sim 5.5$ K.  Figures 1(e) and 1(f) compare the temperature dependence of zero-bias conductance $G(0)$ with the in-plane resistivity $\rho_{xx}$. These plots clearly demonstrate a persistent conductance ``tail'' up to $T_c^{\#}$. This phenomenon of excess conductance spectra has been consistently observed in prior soft point-contact studies on CsV$_3$Sb$_5$ and KV$_3$Sb$_5$ \cite{LuXPRB2021,HeMCPRB2022}.  We emphasize that six separate CsV$_3$Sb$_5$ samples were measured, yielding six distinct sets of Andreev reflection spectra, all of which exhibites nearly identical spectral behaviors.

\begin{figure}
\includegraphics[scale=0.55]{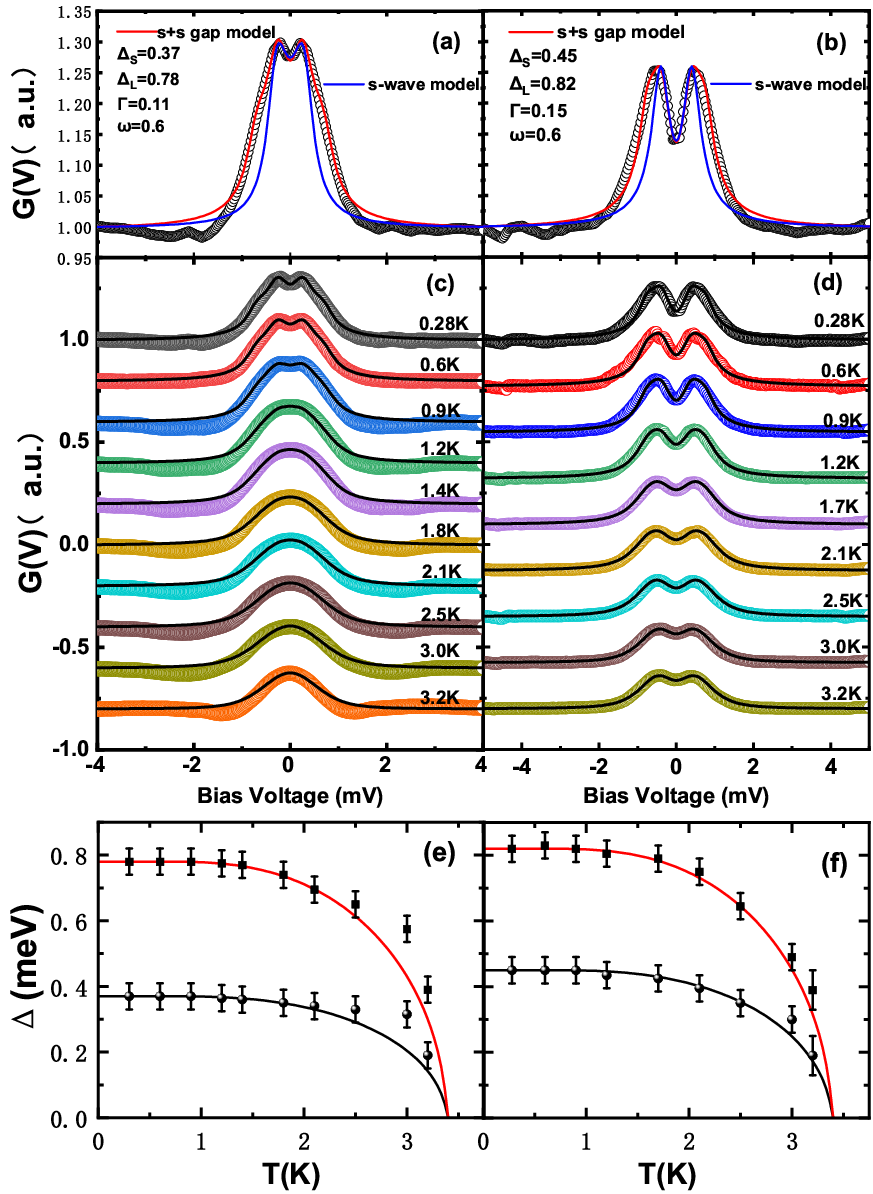}
\caption{(a)(b)The normalized $c$-axis Andreev conductance spectrum $G(V)/G_N$ at the lowest $T=0.28$ K for \#1 and \#2 junctions, respectively. The solid red curves are their BTK fits with a two-isotropic gap model. The solid blue curves are their BTK fits with a single $s$-wave gap model, for comparison.  (c)(d) The normalized $c$-axis Andreev conductance spectra $G(V)/G_N$ at various $T$s from 0.28 K to 3.5 K \#1 (c) and \#2 (d) junctions, respectively.  The solid curves are their BTK fits using a two-isotropic gap model. (e)(f) The corresponding superconducting gap $\Delta(T)$ as a function of $T$, which closely follows the BCS $\Delta-T$ function.}
\label{2}
\end{figure}

To quantitatively assess the gap structure of CsV$_3$Sb$_5$, we examined the PCAR spectroscopy along with the sensing current parallel to both the $c$-axis and $ab$-plane junctions for two different CsV$_3$Sb$_5$ samples (\#1 and \#2).  Fig. 2(a) and 2(b) show the normalized Andreev conductance spectra $G(V)/G_N$ at the lowest $T$ of 0.28 K with sensing current along the $c$-axis ($c$-axis junction) for \#1 and \#2, respectively.  As shown, the $c$-axis junctions clearly show a double-peak features in PCAR spectra, signaling a nodeless gap state.  Based on prior PCAR studies \cite{LuXPRB2021,HeMCPRB2022}, the spectra are more accurately described using a modified Blonder–Tinkham–Klapwijk (BTK) model with two $s$-wave conductance channels $G(V) = \omega G_S(V) + (1 - \omega) G_L(V)$, where $\omega$ quantifies the relative spectral weight of the two channels. Noted we also analyzed such PCAR spectral data of $c$-axis junctions using a single $s$-wave gap model for comparison, as shown in Fig. 2(a) and 2(b).  The two $s$-wave-gap fitting give the best result, yielding a small gap value of $\Delta_S \simeq 0.37$ meV for \#1 and 0.45 meV for \#2, and a large gap value of $\Delta_L \simeq 0.78$ meV (\#1) and 0.82 meV (\#2), with the spectral weight of $\omega \sim 0.60$ for both junctions. The corresponding broadening parameters $\Gamma$ are $\Gamma_S \simeq 0.13$ meV (\#1) and 0.15 meV (\#2), respectively.  The barrier strength, or junction transparency are $Z = 0.54$ and 0.78 for junction \#1 and \#2, respectively. Note that an
isotropic two-gap model cannot produce reasonable results.

The $T$-dependent PCAR spectra and their fits based on the above-obtained parameters are shown in Fig. 3(c) and 3(d) for \#1 and \#2, respectively.  As shown, the two $s$-wave gap BTK model provides good fits to all $G(V)$ curves. In the fitting process, $\omega$ is kept constant for each sample, while $\Gamma$ increases slightly with $T$ up to $T_c$. The resulting gaps $\Delta_L$ and $\Delta_S$ for samples \#1 and \#2 are plotted as a function of $T$, shown in Figs. 2(e) and 2(f), respectively. The obtained gaps can be approximated by an empirical BCS formula: $\Delta(T) = \Delta_0 \tanh(\alpha \sqrt{T_c/T -1})$ where $\alpha$ is adjustable parameter. For samples \#1 and \#2, $\Delta_L \sim 0.80 $ meV, and $\Delta_S\sim 0.41$ meV.  Thus, the coupling strength is estimated as: $2\Delta_L/k_BT_c\simeq 5.47$ for larger superconducting gap in strong coupling limit, and  $2\Delta_S/k_BT_c \simeq 2.73$ for smaller superconducting gap in weak-coupling strength, in align quantitatively with those values of the prototypical two-gap superconductor MgB$_2$ \cite{Szabo2001,Gonnelli2001}, in which the superconducting gaps satisfy the relation $2\Delta_L/k_BT_c > 2\Delta_{\text{BCS}}/k_BT_c (3.52) \geq 2\Delta_S/k_BT_c$ \cite{Iavarone2002}. These values are in good agreement with previously reported results on the similar CsV$_3$Sb$_5$ $c$-axis junctions \cite{LuXPRB2021,HeMCPRB2022}.

On the other aspect, the point-contact spectra with sensing current $I$ along the $ab$-plane ($ab$-plane junction) shown in Figs. 3(a) and 3(b) display a single broad zero-bias conductance peak in the gap region, characteristic of an anisotropic gap. These spectra are fitted to an anisotropic $s$-wave BTK model. Considering the six-fold symmetry inherent to the Kagome lattice and insights from ARPES measurements, we model the anisotropic $s$-wave superconducting gap using the form $\Delta^{a} \times [(1 - r) + r \cos^2(3\theta)]$ \cite{Feng2025} or the equivalently $\Delta^a \times [(1-r)+r \cos(6\theta)]$ \cite{elecrad}. In this gap function, $r$ represents the gap anisotropy ratio, varying from $r = 0$ (isotropic $s$-wave state) to $r$ = 1 (completely nodal gap state). As shown, the anisotropic gap BTK model provides an excellent description of the normalized $G(V)$ at the lowest $T$. The fits yield $\Delta^a \simeq 0.50$ meV for \#1 crystal and 0.49 meV for \#2 crystal, and $\Gamma$ = 0.18 meV for both junctions. The gap anisotropy ratio $r=0.70\pm 0.10$ is resolved for the anisotropic gap of the $ab$-plane junctions.

\begin{figure}
\includegraphics[scale=0.55]{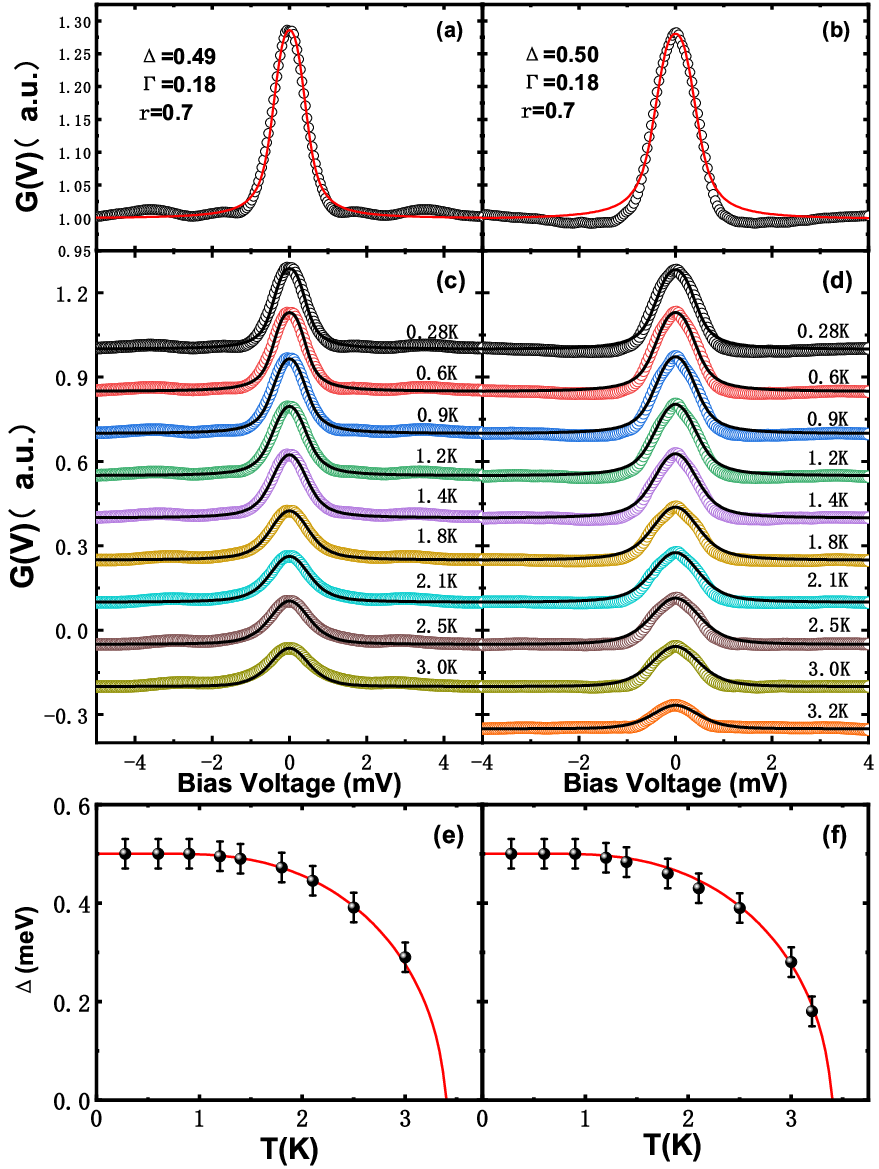}
  \caption{(a)(b)The normalized $ab$-plane Andreev conductance spectrum $G(V)/G_N$ at the lowest $T=0.28$ K for \#1 and \#2 junctions, respectively. The solid red curves are their BTK fits with an anisotropic gap model. (c)(d) The normalized $ab$-plane Andreev conductance spectra $G(V)/G_N$ at various $T$s from 0.28 K to 3.5 K \#1 (c) and \#2 (d) junctions, respectively.  The solid curves are their BTK fits using an anisotropic gap model. All the conductance spectra and their corresponding fitting curves are shifted downward for clarity, excepted the top ones. (e)(f) The corresponding superconducting gap $\Delta(T)$ as a function of $T$, which closely follows the BCS $\Delta-T$ function.}
\label{2}
\end{figure}

Figures 3(c) and 3(d) show the normalized point-contact conductance spectra of the $ab$-plane junction, measured over a temperature range from 0.28 K to 3.4 K. The experimental data were fitted to the anisotropic-gapped BTK model. In the fitting process, the anisotropy $r = 0.70$ keep constant. The temperature evolution of the anisotropic gap extracted from the fits is presented in Fig.3(e) and 3(f) for \#1 and \#2, respectively. The extracted SC gap $\Delta^a$ obeys the BCS gap-$T$ formula with the anisotropic SC gap $\Delta^a\simeq 0.50$ meV in the superconducting ground state.
Based on these obtained values, there exists a gap minimum $\Delta^a_{mini}=\Delta^a(1-r)\simeq 0.15$ meV.  Recently, a prominent ARPES experiment directly measure the SC gap in the momentum space in CsV$_3$Sb$_5$ single crystals.  It is shown that the SC gaps in the $\alpha$-band Fermi surface which from Sb-5$p$ orbital and $\delta$-band Fermi surface from V-3$d$ orbital are isotropic, whereas the SC gap in the $\beta$ Fermi surface from induced V-$3d$ orbital is highly anisotropic with a gap mimimum of 0.10 meV \cite{lateARPES2024}. Such a nodal $s$-wave gap, without sign reversal of SC order parameter, is completely consistent with our soft PCAR results.

In final, our Andreev reflection spectroscopy measurements reveal a significant gap anisotropy in CsV$_3$Sb$_5$, quantified at approximately 70\% with a gap mimimum 0.15 meV. However, given the known strong suppression of gap anisotropy by disorder/impurity effect in kagome superconductors, and the limited sample purity (indicated by the residual resistivity ratio, RRR), this measured value likely represents a lower bound. We posit that in intrinsically clean CsV$_3$Sb$_5$ samples, the gap anisotropy could approach 100\%, implying possibly a true node gap.

\emph{4. Conclusion:} A directional PCAR experiment was performed to probe the SC gap structure of kagome superconductor CsV$_3$Sb$_5$.  The point-contact Andreev conductance spectra feature a strongly anisotropic gapping behavior, signalling the existence of an anisotropic gap in addition to isotropic SC gaps.  Quantitative analysis of the spectral data estimates a gap-anisotropy $\sim$70\% and the gap minimum $\sim 0.15$ meV.  Our direct observation of the orbital-selective anisotropic Cooper pairing in pristine CsV$_3$Sb$_5$ points to the unconventional pairing mechanisms and could also be consistent with the SC gap structure of a $2\times2$ pair density wave, providing a foundation for further understanding the intertwinement with other electronic ordering states, such as the time-reversal symmetry breaking, loop current ordering and electronic nematicity.

\end{document}